\documentclass[12pt]{iopart}
\usepackage{iopams,graphicx,epsfig}
\begin{document}

\title[Nonlinear susceptibility]{
Non-linear susceptibilities of spherical models
}

\author{Leticia F. Cugliandolo} 
\address{Universit\'e Pierre et Marie Curie -- Paris VI, 
Laboratoire de Physique
Th{\'e}orique et Hautes {\'E}nergies, UMR 7589,
5\`eme {\'e}tage, Tour
25, 4 Place Jussieu, 75252 Paris Cedex 05, France 
}
\ead{leticia@lpt.ens.fr}
\author{David S. Dean}
\address{
Laboratoire de Physique Th\'eorique, CNRS UMR 5152,
IRSAMC, Universit\'e Paul Sabatier, 118 route de Narbonne, 31062
Toulouse Cedex 04, France}
\ead{dean@irsamc.ups-tlse.fr}
\author{Hajime Yoshino}
\address{
Department of Earth and Space Science, Faculty of Science, 
Osaka University, Toyonaka 560-0043, Japan}
\ead{yoshino@ess.sci.osaka-u.ac.jp}

\date{27th October 2006}

\begin{abstract}
The static and dynamic susceptibilities for a general class of
mean field random orthogonal spherical spin glass models are studied.
We show how the static and dynamical properties of the linear and nonlinear
susceptibilities depend on the behaviour of the density of states of the two
body interaction matrix in the neighbourhood of the largest eigenvalue.
Our results are compared with experimental results and also with 
those of the droplet theory of spin glasses.
\end{abstract}  
\pacs{05.70.Fh, 75.10.Nr, 75.50.Lk}
\maketitle

\section{Introduction and summary of results}

Static non-linear susceptibilities serve to characterise a
phase transition and its universality class, especially
in spin-glass systems in which the linear counterpart 
does not show a divergence but just a cusp at a finite
critical temperature. Experimental results for 
standard spin-glass samples have now been available for over 
20 years~\cite{Leog,Levy} while studies of 
more exotic systems exhibiting spin-glass ordering, 
such as manganites, are currently being carried out~\cite{Nair}.
  
Recently it has been pointed out that dynamic non-linear susceptibilities can
also be very useful to characterise, and eventually  
understand, the slow dynamics of super-cooled liquids,
their arrest~\cite{Silvio,Bobi,Cristina,long},
and the non-equilibrium dynamics of the low temperature regime~\cite{us,Mayer}.

In order to separate the truly non-trivial behaviour of glassy
systems (with or without quenched disorder) from phenomena also
present in simpler cases, such as ferromagnetic domain growth and
other simple phase ordering mechanisms, it is important to understand
the behaviour of the static and dynamic generalised susceptibilities
in solvable toy models.  The aim of this paper is to present a
detailed analysis of static and dynamic linear and non-linear magnetic
susceptibilities in {\it generic spherical disordered models with
two-body interactions}~\cite{spher}-\cite{Becuig}.  
We shall investigate how
these quantities depend on the density of states of the two-body
interaction matrix.  In the study of susceptibilities, {\em i.e.} the
influence of an external field $h$ on the system, two thermodynamic
limits in $h$ and $N$ (the number of spins or system size) can be
considered,
\begin{enumerate}

\item
The applied field goes to zero first and the thermodynamic limit 
is taken after:
\begin{displaymath}
\lim_{N\to\infty} \; \lim_{h\to 0} 
\; . 
\end{displaymath}

This is the limit which always coincides with static  fluctuation
dissipation relations relating susceptibilities to correlation functions.
\item
The field goes to zero after the thermodynamic limit 
is taken:
\begin{displaymath}
\lim_{h\to 0} \; \lim_{N\to\infty} 
\; . 
\end{displaymath}
This is the limit of the critical isotherm.

\end{enumerate}

It will turn out that these two limits are equivalent in the high
temperature phase, however they differ in the low temperature regime.

In our study we do not consider finite $N$ corrections to the density
of states $\rho$ of the interaction matrix which is taken to be
deterministic and does not vary from sample to sample. The underlying
phase transition in these models is similar to Bose Einstein
condensation where at low temperatures the system develops a
macroscopic condensation onto the eigenvector corresponding to the
largest eigenvalue $\mu_0$ of the interaction matrix. In other words
the system develops a macroscopic magnetisation in the direction of
this eigenvector. When the largest eigenvalue of the interaction
matrix is bounded these models can present a finite temperature
continuous phase transition~\cite{KTJ}-\cite{espagnoles}.  Whether or not this
transition is realised depends on how the density of states
$\rho$ vanishes at $\mu_0$. In cases with a density of states with
tails that extend to infinity there is no finite $T_c$, see {\it
e.g.}~\cite{SCM,Secu}. Here we restrict out attention to density
of states where the maximal eigenvalue is bounded.  In the high $T$
phase the system is paramagnetic (or liquid) while, as mentioned
above, in the low $T$ regime it is ordered via a Bose-Einstein
condensation mechanism.  A finite magnetic field may or may not kill
the static transition depending on the decay of the density of states
close to the edge at $\mu_0$. If the density of states behaves as
$\rho(\mu) \sim (\mu_0-\mu)^\alpha$ about the edge at $\mu_0$ there is
only a finite temperature transition if $\alpha >0 $. In the presence
of an external field, this transition is killed if $\alpha <1$ but
there is still a transition if $\alpha > 1$. In much of our study we
will concentrate on the regime where $\alpha\in (0,1)$. The divergence
of all the susceptibilities studied here, and other critical exponents,
can all be expressed in terms of the exponent $\alpha$.

From the point of view of the dynamics of these models, a typical
initial condition does not have a macroscopic overlap with the
condensed low temperature equilibrium configurations. Consequently the
low temperature dynamics is slow and the equilibrium condensation is
not reached in finite times with respect to $N$.  The relaxation
occurs out of equilibrium and correlation and linear response
functions age with similar scaling forms to ferromagnetic domain
growth~\cite{Ciuchi,cude}. In the case where a finite
magnetic field is applied, and when $\alpha <1$ i.e. the field kills 
the static phase transition,  a
charactersitic time-scale $t^*(h)$ is introcuced that separates a 
transient non-equilibrium regime from the final approach to equilibrium in the 
disordered phase~\cite{Cude2}.

\subsection{Summary of results}

We show that in case (i) the static linear susceptibility does not
have a cusp at $T_c$ and obeys a Curie law down to $T=0$. The first
static non-linear susceptibility ($\chi_3$) however diverges linearly
with the size of the system at all $T<T_c$. In the limit (ii) the static linear
susceptibility exhibits a cusp and the behaviour of the static
non-linear susceptibility depends explicitly on the decay of the
density of states of the interaction matrix at its upper edge. For a
density of states decaying as the power law $(\mu-\mu_0)^\alpha$, $\chi_3$
diverges if $\alpha<1/2$, vanishes if $\alpha>1/2$ and takes a
finite value if $\alpha=1/2$.  

In a dynamic study, using a Langevin stochastic evolution for the
continuous spins, we elucidate the approach of the linear and first
non-linear susceptibilities to their asymptotic static limits in the
more interesting low-temperature phase. In short, we find that using
the first order of limits the linear susceptibility, $\chi_1$, approaches its
asymptotic static value with an exponential decay of characteristic
time $\tau = N(\beta-\beta_c)$. In the second case $\chi_1$ achieves
its asymptotic value with a power law decay $t^{-\alpha}$ but the
non-linear susceptibility $\chi_3$ diverges or vanishes, depending on
$\alpha$ being larger or smaller than $1/2$, with a power law
$t^{1-2\alpha}$ - in agreement with the static calculation. In the
case (i), where the zero field limit is taken first, $\chi_3$
approaches its asymptotic value exponentially with the same
characteristic time as the linear susceptibility, $\tau=N(\beta-\beta_c)$.

We stress the fact that the distinction between the two limiting procedures
(i) and (ii)   is especially important in
numerical simulations (and experiments). If one takes the first order
of limits no special feature at $T_c$ is observed in the
linear susceptibility.  One needs to use a sufficiently large field and fall
into the second case, to see the critical behaviour of the linear
susceptibility.

The difference between these two limiting procedures has been
discussed in a number of papers. Within the droplet model Fisher and
Huse derive scaling forms for the susceptibility in both
cases~\cite{Fihu}. More recently, Yoshino and Rizzo~\cite{Yori}
studied the linear response in mesoscopic disordered models with one
step replica symmetry breaking using the Thouless-Anderson-Palmer
approach and found stepwise signals (as anticipated by Kirkpatrick and
Young~\cite{Kiyo} and by Young, Bray, and  Moore~\cite{Brmoyo} in their
studies of the Sherrington-Kirkpatrick spin-glass). They also discussed the 
importance of considering the two limiting procedures (i) and (ii). 

In experiments and numerical simulations, linear and non-linear
susceptibilities are usually obtained as functions of (higher order)
correlations functions by virtue of the fluctuation-dissipation
relations (in equilibrium) or its extensions (out of equilibrium).  Some
fluctuation-dissipation relations linking non-linear susceptibilities
and correlation functions in the absence of a field, in glassy systems
out of equilibrium, can be found in~\cite{SCM,Monthus}.  Numerical
recipes to compute the linear susceptibility using fluctuation
dissipation relations, also valid  out of equilibrium, have been proposed
by a number of authors~\cite{Chatelain,Fede,Corberi}.  All these
expressions correspond to  the order of limits (i).

Having presented our main results above, the body of the paper presents
technical details of their derivation.  In Sect.~(\ref{sec:model}) we
introduce the model and the notation. In Sect.~(\ref{sec:statics}) we
present the free-energy density and the main observable we shall
study, the magnetisation density. Sect.~(\ref{sec:langevin}) is devoted
to the analysis of the Langevin dynamics and the derivation of the
dynamic susceptibilities. Finally in Sect.~(\ref{sec:discussion})
we further discuss our results and present our conclusions. 

\section{The model}
\label{sec:model}

The Hamiltonian  considered is  for a system $N$ continuous spins 
\begin{displaymath}
H = -{1\over 2}\sum_{ij} J_{ij}S_i S_j - \sum_i h_i S_i
\; ,
\end{displaymath}
with the spherical constraint $\sum_i S_i^2 = N$. We denote the fixed, or
deterministic, density of
states of the interaction matrix $J$ by
\begin{displaymath}
\rho(\mu) = {1\over N}\sum_\lambda \delta(\mu -\lambda).
\end{displaymath}
In what follows we consider the generalised random orthogonal class of
models with
\begin{displaymath}
J \equiv {\cal O}^T J {\cal O}
\end{displaymath}
where $\equiv$ indicates statistically identical and ${\cal O}$ is 
a random rotation chosen with the Haar measure. 
The behaviour of the statics of this class of generalised random orthogonal 
models with  Ising spins has been studied in \cite{cdl} and the first 
such models were studied in \cite{rom}.

We assume that the density 
of states $\rho(\mu)$ has an edge at $\mu_0$ and 
we use density of states $\rho$ that admit the following expansion 
about $\mu = \mu_0$
\begin{equation}
\rho(\mu) = \sum_{n=0}^{\infty}c_n (\mu_0-\mu)^{\alpha +n}
\label{eqexrho}
\; .
\end{equation}
For the integrability of $\rho(\mu)$ we must have that $\alpha > -1$.

We pay special attention to the 
The Gaussian ensemble 
where the elements of the (symmetric) $J_{ij}$ are Gaussian of zero mean 
and variance $1/N$. In this case  we find 
the Wigner Semi-Circle Law for the density of states
\begin{displaymath}
\rho(\mu) = {1\over 2 \pi}\sqrt{4 - \mu^2} \label{eqwig}
\; , 
\end{displaymath}
clearly $\alpha=1/2$, $\mu_0=2$ and $c_0= 1/\pi $.  
This model is 
statistically invariant with respect to such random rotations or 
transformations. However it does not strictly belong to this class
of models, as the density of states fluctuates from sample to sample. However
these sample to sample fluctuations are unimportant for the computation
of extensive thermodynamics quantities.  
This model is then equivalent, in the thermodynamic limit to the spherical 
Sherrington-Kirkpatrick disordered model studied in a
number of 
publications~\cite{KTJ}-\cite{Becuig}.

The spherical ferromagnet in $d$ dimensions, and its continuum 
version, the O(N) model,  can also be included in this 
family of models but we shall not discuss them in detail here. 
From the calculational point of view it is practical to consider an applied  
external field ${\bf h}$  of the form 
\begin{displaymath}
h_i = h \sigma_i
\; , 
\end{displaymath}
where $\sigma_i$ are independent Gaussian random variables of mean zero and
variance one. This means that the distribution of ${\bf h}$ is also invariant
under random rotations, {\em i.e.}
\begin{displaymath}
{\bf h} \equiv {\cal O}{\bf h}
\; .
\end{displaymath}
Let us note that ${\overline {\bf h^2}} = h^2 N$ where the over-line
indicates averaging over realisations of the field ${\bf
h}$. Consequently the central limit theorem tells us that for large $N$
we have ${\bf h}^2 = h^2 N +O(\sqrt{N})$, this means that after a
suitable rotation $h_i = h + O(1/\sqrt{N})$, {\em i.e.} up to
$1/\sqrt{N}$ corrections, there is a basis where ${\bf h}$ is a
uniform vector.  The magnetisation induced in the direction of the
field ${\bf h}$ is given by
\begin{displaymath}
m = {1\over N}\sum_i \langle \, S_i \sigma_i \, \rangle
\end{displaymath}
and for the reasons given above, in the large $N$ limit $m$ will be a function
of $h$. 

\section{The statics}
\label{sec:statics}

\subsection{The free-energy density}
\label{subsec:freeenergy}

Standard arguments 
lead to a variational expression for the 
free-energy (up to constant terms irrelevant for our calculations) per spin~\cite{KTJ},
\begin{displaymath}
\beta f(z) = {1\over 2}\int d\mu \ \rho(\mu) \ln\left(z-\mu\right)
-{\beta h^2\over 2}\int d\mu \; {\rho(\mu)\over z-\mu} - {\beta z \over 2}
\; , 
\end{displaymath}
where $z$ is the Lagrange multiplier introduced to impose the spherical
constraint. The free-energy per site $f$ is obtained {\it via} a 
saddle-point calculation over  $z$ as
\begin{displaymath}
 f = {\rm min}_z \ f(z)
\; , 
\end{displaymath}
and the corresponding saddle point equation is 
\begin{displaymath}
\langle\langle\,  {1\over z-\mu}\, \rangle\rangle + \beta h^2 
\; \langle\langle\, 
{1\over (z-\mu)^2}\, \rangle\rangle = \beta \label{eqsp}.
\end{displaymath}
Here we have used the short hand notation
\begin{displaymath}
\langle\langle \, \, \ldots \, \rangle\rangle_{\mu} \equiv \lim_{N \to \infty} 
\frac{1}{N}\sum_{\mu} \ldots 
\end{displaymath}
to represent the average over eigenvalues. 

\subsection{The phase transition}

Whether or not this model exhibits a phase transition depends on the
density of states $\rho(\mu)$ of the interaction matrix $J_{ij}$.
The constraint equation may be written as 
\begin{equation}
F(z,\beta,h) = \beta 
\label{eqcons}
\end{equation}
with
\begin{equation}
F(z,\beta,h) = 
\langle\langle \; {1\over z-\mu} \; \rangle\rangle 
+ 
\beta h^2 \;\;
\langle\langle \; {1\over (z-\mu)^2} \; \rangle\rangle 
\; . 
\label{eqF}
\end{equation}
In this analysis one is required to choose a solution 
$z \geq \mu_0$ where $\mu_0$ is the largest eigenvalue of the 
interaction matrix. If $F(z,\beta,\mu)$ diverges as $z \to \mu_0$ then the
constraint equation can always be satisfied in a continuous manner and there
will be no finite temperature phase transition. 

In the case $h=0$ there is a transition if $\alpha >0$.
In the case $h \neq 0$ there is a transition if $\alpha > 1$. Thus
the presence of a finite field for $\alpha \in (0,1)$ kills the transition.

For models where $\alpha \in (0,1)$ the field has a very singular effect 
on the thermodynamics. In this range of $\alpha$ and in the 
absence of a field there is a critical  temperature $T_c = 1/\beta_c$ 
defined by
\begin{equation}
\beta_c = \langle\langle \; {1\over \mu_0 - \mu} \; \rangle\rangle
\; .
\label{eqbc}
\end{equation}
For $\beta <\beta_c$, the Lagrange multiplier at zero field $z_0$ 
varies continuously with $\beta$ and is the solution to  
\begin{equation}
\langle\langle \; {1\over z_0- \mu} \; \rangle\rangle = \beta
\; ,
\label{eq:lagrange}
\end{equation}  
whereas for $\beta > \beta_c$ we have $z_0 = \mu_0$ up to $1/N$ corrections
and the spherical 
constraint is satisfied by a macroscopic condensation onto the eigenvector
corresponding to the eigenvalue $\mu_0$ in a manner analogous to 
Bose-Einstein condensation. More specifically, we decompose $\langle\langle 
\dots \rangle\rangle$ into two parts, the first being over the 
largest eigenvalue $\mu_0$ and the second being over the remaining
eigenvalues, which can be treated as a continuum, which we denote 
by $\langle \langle \dots \rangle\rangle_c$. In this notation the 
constraint equation becomes
\begin{displaymath}
\frac1{N} \frac{1}{z_0-\mu_0} + 
 \langle \langle \; \frac1{z_0-\mu} \; \rangle\rangle_c =\beta
\; . 
\end{displaymath}
To leading order in $1/N$ the solution to this equation is 
\begin{equation}
z_0 = \mu_0 + \frac1{N} \, \frac{1}{(\beta-\beta_c)}
\; . 
\label{eq:z0-Ncorr}
\end{equation}

\subsection{The susceptibilities}
\label{sec:susceptibilites}

The magnetisation in the direction of the applied field is given by
\begin{equation}
m(h) = -{\partial f\over \partial h} =  h \; 
\langle\langle \; \frac1{z-\mu} \; \rangle\rangle
\; .
\label{eq:magn-dir-field}
\end{equation}
In order to fully determine the magnetisation as a function of the
applied field strength we need to know how $z$ varies as a function of
$h$. From the form of the free-energy we see that $z$ must be an even
function of $h$ and thus
\begin{equation}
z(h) = \sum_{n=0}^\infty z_n h^{2n}
\;,
\label{eqexz}
\end{equation}
at least at high temperatures.
The magnetisation then has an expansion in terms of the 
generalised susceptibilities 
\begin{equation}
m(h) =  \sum_{n=0}^{\infty} {\chi_{2n+1}\over (2n+1)!} h^{2n+1}
\;  
\label{eq:chi-def2}
\end{equation}
or, equivalently, the $n$-th order susceptibility is just
\begin{equation}
\chi_n = 
\left. \frac{\partial^n m(h)}{\partial h^n} \right|_{h=0}
= - \left. {\partial^{n+1} f\over \partial h^{n+1}} \right|_{h=0}
\; . 
\label{eq:chi-def}
\end{equation}

The cases $\beta > \beta_c$ must be considered separately as
the starting point for the expansion, the value of $z_0$
in equation~(\ref{eqexz}), is different.
In both cases we expand equation~(\ref{eqcons}) in powers of $h$ using the 
expansion of equation~(\ref{eqexz}) for $z$.

\subsubsection{High temperatures, $T>T_c(h=0)$.}
\label{subsec:highT}

First we consider the high temperature regime $\beta < \beta_c$.
Using the variables  $g_n$ defined by 
\begin{equation}
g_n = \langle\langle \; {1\over (z_0-\mu)^n} \; \rangle\rangle
\; ,
\label{eq:chi1}
\end{equation}
we find that
\begin{eqnarray}
z(h) &=& 
z_0 + \beta \; h^2 -\beta^2 {g_3\over g_2} \; h^4 + 
2 \beta^3{g_4\over g_2} \; h^6
\nonumber\\ 
&&
+\beta^4{(g_3^3 -3 g_2 g_3 g_4 - 3g_2^2 g_5)\over g_2^3} \; h^8 
+
O(h^{10})
\end{eqnarray}
and 
\begin{eqnarray}
m(h) &=& 
\beta \; h -\beta g_2 \; h^3 + 2\beta^2 g_3 \; h^5 - 
{\beta^3\left( 
2g_3^2 + 3 g_2 g_4\right)\over g_2} \; h^7 
\nonumber\\
&& 
+
2 \beta^4 
{\left(5 g_3 g_4 + 2 g_2 g_5
\right)\over g_2} \; h^9 +O(h^{11})
\; . 
\label{eq:expansion-highT}
\end{eqnarray}
Note that from the zero field constraint equation for $z_0$ one has
that $g_0 = \beta$. The first three terms look particularly simple,
suggesting a simple pattern which however breaks down at the next
order.  From this expression one can read the $n$-th order
susceptibility $\chi_n$, see equations~(\ref{eq:chi-def2}) and
(\ref{eq:chi-def}), $\chi_1=\beta$, $\chi_3=-6\beta g_2$ and so on and
so forth. The linear susceptibility has a Curie-Weiss behaviour all
the way up to $T=T_c$. As one approaches $T_c$ the Lagrange multiplier $z$
approaches $\mu_0$ and so we write 
\begin{equation}
z\sim \mu_0 + \delta z
\label{z-closeTc}
\end{equation}
where  $\delta z$ is small with respect to $\mu_0$. 

The integrals in the definition of the parameters $g_n$, with $n\geq 1$
are dominated by the divergence of the denominator at the edge 
$\mu\sim \mu_0$ for all $\alpha < 1$ and therefore 
\begin{displaymath}
g_n \approx c_0 \int_0^\infty  d\epsilon \; 
\frac{\epsilon^\alpha}{(\epsilon+\delta z)^n } 
= c_0 (\delta z)^{\alpha+1-n} \int_0^\infty  du \; 
\frac{u^\alpha}{(1+u)^n } 
\; . 
\end{displaymath}
The scaling of $\delta z$ with $T-T_c$ is obtained from the analysis 
of equation~(\ref{eq:lagrange}). Introducing (\ref{z-closeTc}) in 
(\ref{eq:lagrange}) one finds
\begin{equation}
\langle\langle 
{\delta z\over (\mu_0 +\delta z - \mu)(\mu_0-\mu)}
\rangle\rangle = \beta_c -\beta
\end{equation}
which for $\delta z$ small gives
\begin{equation}
c_0 I(\alpha) (\delta z)^{\alpha} \approx (\beta_c -\beta)
\end{equation}
where 
\begin{equation}
I(\alpha) = \int_0^{\infty} du\ {u^{\alpha-1}\over 1 + u} = {\pi\over 
\sin\left[\pi (1-\alpha)\right]}
\; . 
\label{ialpha}
\end{equation}
This thus yields 
\begin{equation}
\delta z \approx   
\left( {\beta -\beta_c \over c_0 I(\alpha)}\right)^{1\over \alpha}\sim (T -T_c)^{1/\alpha}
\end{equation}
and consequently 
\begin{equation}
g_n \approx c_0\int_0^\infty  du \; 
\frac{u^\alpha}{(1+u)^n } \left[{\beta -\beta_c\over c_0 I(\alpha)}\right]^{1+(1-n)/\alpha} \sim (T-T_c)^{1+(1-n)/\alpha}
\; . 
\label{eq:gn-asymptotic}
\end{equation}

From this we find that close to $T_c$, all
the non-linear susceptibilities behave as 
\begin{displaymath}
\chi_n \sim (T-T_c)^{1+\frac12 (1-n)/\alpha}
\; . 
\end{displaymath}
This expression diverges as soon as $n>1+2\alpha$. This means that for
$\alpha \in (0,1)$ all $\chi_n$ with $n\geq 3$ diverge.  Note that the
order of limits (i) $N\to \infty \; h\to 0 $ or (ii) $h\to 0 \;
N\to\infty$ is irrelevant at high temperatures.

\subsubsection{Low temperatures, $T<T_c(h=0)$.}
\label{subsec:lowT}

At low temperatures we have to consider separately the 
two thermodynamic limits mentioned in the introduction. 

\vspace{0.25cm}
\noindent
{(i) $N\to\infty \;\; h\to 0$}
\vspace{0.25cm}

In this case the solution to the constraint equation for $z_0$ is
given by equation~(\ref{eq:z0-Ncorr}) therefore $z_0$ is always strictly
greater than $\mu_0$ and the expansion given by
equation~(\ref{eq:expansion-highT}) is still valid as the expressions
for the $g_n$ are still finite and we find that
\begin{eqnarray}
g_n &\approx & N^{n-1}(\beta -\beta_c)^n + \langle\langle
{1\over \mu_0 +{1\over N(\beta-\beta_c) } -\mu}\rangle\rangle_c \nonumber
\\ &\approx& N^{n-1} (\beta -\beta_c)^n + 
O\left( (N(\beta-\beta_c))^{n-1-\alpha}\right)
\end{eqnarray}  
As an example, the
first non-linear susceptibility is given by 
\begin{equation}
\chi_3 = -6\beta (\beta-\beta_c)^2 \;  N + O(N^{1-\alpha}) 
\; . 
\label{eq:static-value}
\end{equation} 

\vspace{0.25cm}
\noindent
{(ii) $h\to 0 \;\;  N\to\infty$}
\vspace{0.25cm}

When the thermodynamic limit is taken in the 
presence of an applied field, 
the expansion for $\beta > \beta_c$ is carried out with $z_0 = \mu_0$
but we do not assume an analytic expansion for $z$ and thus we write
\begin{displaymath}
z = \mu_0 + s(h)
\; ,
\end{displaymath}
where $s(h) \to 0 $ as $h\to 0$.
The constraint equation now reads
\begin{equation}
\langle \langle \, 
{s(h)\over (\mu_0-\mu)[\mu_0-\mu + s(h)]}\, \rangle\rangle
- \beta h^2 
\langle\langle \, {1\over [\mu_0-\mu + s(h)]^2}\, \rangle\rangle
= \beta_c -\beta 
\label{eqf1}
\; .
\end{equation}
Expanding equation~(\ref{eqf1}) to leading order in $s(h)$ and examining the limit
$h\to 0$ we see that to lowest  order in $h$ the function $s(h)$ is given by  
\begin{displaymath}
s(h)= 
h^{2\over 1-\alpha} 
\; 
\left[ {\beta c_0\alpha 
I(\alpha)\over (\beta-\beta_c)}\right]^{1\over 1-\alpha},
\end{displaymath}
where $c_0$ is defined by the expansion of the density of states
equation~(\ref{eqexrho}) about $\mu_0$.
The key point in this derivation is that the most divergent
contribution to the integrals in equation~(\ref{eqf1}) as $s\to 0$ come
from the first term in the expansion of $\rho$ about $\mu_0$. This
most divergent term may be extracted as follows. 
If $\mu_*$ is the minimal eigenvector of $J$, a typical term to evaluate is
\begin{eqnarray*}
&& 
\int_{\mu_*}^{\mu_0} d\mu \;
{(\mu_0-\mu)^{\alpha-1}\over
(\mu_0 -\mu +s)} 
= 
s^{\alpha-1}\int_0^{{\mu_0 -\mu_*}\over s}du \ {u^{\alpha -1}
\over 1+ u}
\approx  
s^{\alpha-1}\int_0^{\infty}du \ {u^{\alpha -1}
\over 1+u} 
\end{eqnarray*}
as $s\to 0$.  Substituting this result into the expression for the
density of states we find that to the first two lowest order terms in
the expansion
\begin{equation}
m(h) = \beta_c h - c_0 I(\alpha) 
\left[ {\beta c_0\alpha I(\alpha)\over 
(\beta-\beta_c)}\right]^{\alpha\over 1-\alpha} \;
h^{1+\alpha\over 1-\alpha}
\; . 
\label{eq:mh-lowT}
\end{equation}

The linear susceptibility is $\chi_1=\beta_c$ leading to the appearance
of a cusp at $T_c$ (where $\chi_1$ passes from the Curie-Weiss law to
becoming a constant).

Interestingly the case where $\chi_3$ exists for $T<T_c$ corresponds
to the case where ${1+\alpha\over 1-\alpha} = 3$ {\em i.e.} $\alpha =
1/2$ which is the Gaussian spherical SK
model~\cite{KTJ}-\cite{Becuig}. 
In the case where $\alpha < 1/2$ we
see that $\chi_3$ is infinite for $T < T_c$, but if $\alpha > 1/2$ it
vanishes, and so $\chi_3 =0$. The spherical SK model thus turns out to be the
marginal case for the non-linear susceptibility $\chi_3$. 
We find that for $T<T_c$ the first non-linear susceptibility behaves as
\begin{equation}
\chi_3 = 
-3 \pi^2 c_0^2 \left({\beta\over \beta -\beta_c}\right)\label{eqchi3+}
\label{eq:Gaussian-lowT}
\end{equation}
while approaching $T_c$ from the 
high temperature region   $\chi_3$ diverges as
\begin{displaymath}
\chi_3 = 
- 6\beta \alpha c_0 I(\alpha) \left(c_0 I(\alpha)\over \beta_c -\beta\right) 
=  -3 \pi^2 c_0^2\left({\beta\over \beta_c -\beta}\right),
\end{displaymath}
which along with equation~(\ref{eqchi3+}) means that in the critical region 
we may write
\begin{displaymath}
\chi_3 = -3 \pi^2 c_0^2 \left|{\beta\over \beta_c -\beta} \right|
\; ,
\end{displaymath}
{\em i.e} the coefficients of the power law divergence in $\chi_3$ are the 
same on each side of the transition.

The Gaussian spherical SK model can be treated 
non perturbatively and solved directly as we show in the next Subsection.

\subsection{The Gaussian case}
\label{eq:subsec:Gaussian}

Let us focus here one the case $\alpha=1/2$ in which the interaction
matrix has Gaussian distributed elements and the density states is
given by the Wigner Semi-Circle law, equation~({\ref{eqwig}). This is 
the spherical SK model. In this case $T_c=J$. 
The integrals over $\mu$ in the constraint equation can now be carried out
explicitly yielding 
\begin{displaymath}
z - \sqrt{z^2 - 4} + 
\frac{\beta h^2}{2} \left( -1 + \frac{z}{\sqrt{z^2-4}}\right)
= 2\beta
\; ,
\end{displaymath}
from where one easily obtains $z(T,h)$ for all values of the
parameters $T$ and $h$ with no need to use a perturbative expansion.
The numerical representation of the solution as a function of $h$ for
three values of the temperature, $T=0.5, \, 1, \, 1.5$, is displayed
in Fig.~\ref{fig1}-left (with solid lines).  These results are
compared to the first order terms in the series expansion valid for
$h\to 0$ in Fig.~\ref{fig1}-right (with thin lines). We see that the
high ($T=1.5$), critical ($T=1$) and low ($T=0.5$) temperature 
curves are indeed very close to the $h^2$ (high and critical $T$s)
and $h^4$ (low $T$) predictions of 
Sects.~\ref{subsec:highT} and~\ref{subsec:lowT}
when $h$ is relatively small.

\begin{figure}[!tbp]
\begin{flushright}
\includegraphics[width=6.5cm,clip=true]{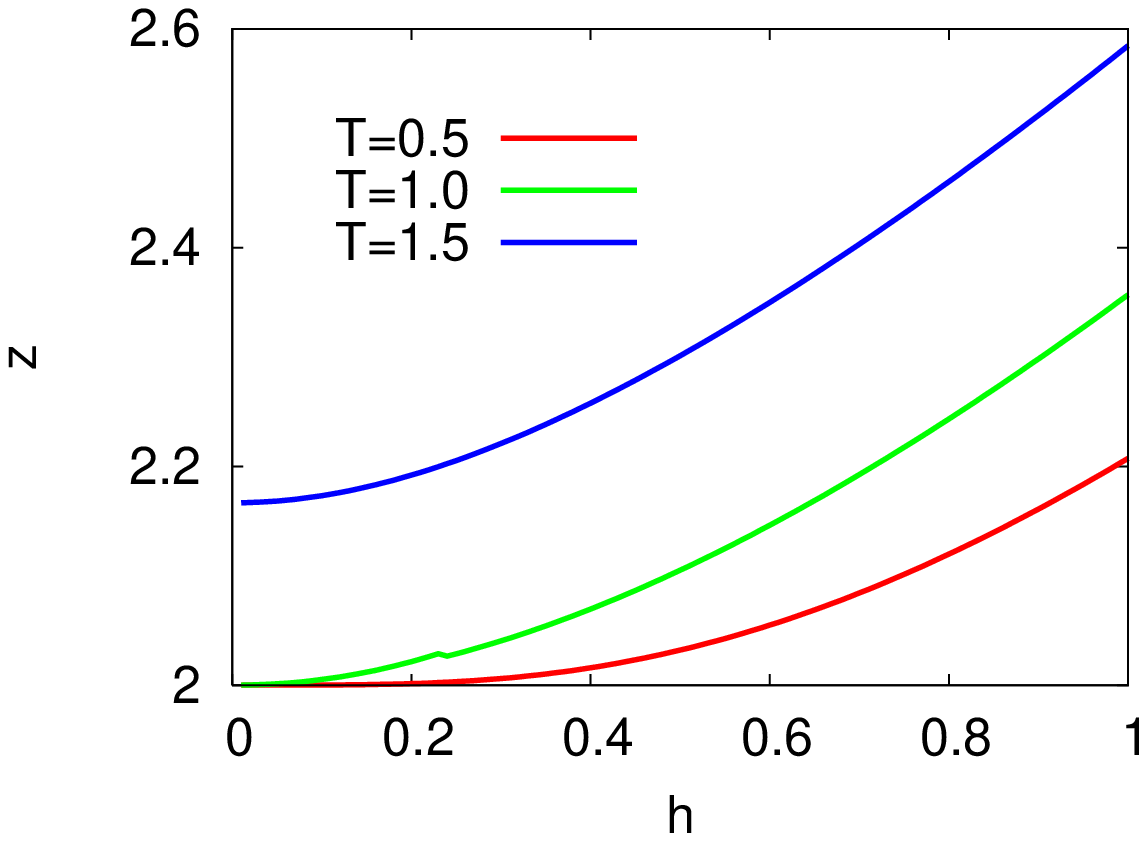}
\includegraphics[width=6.5cm,clip=true]{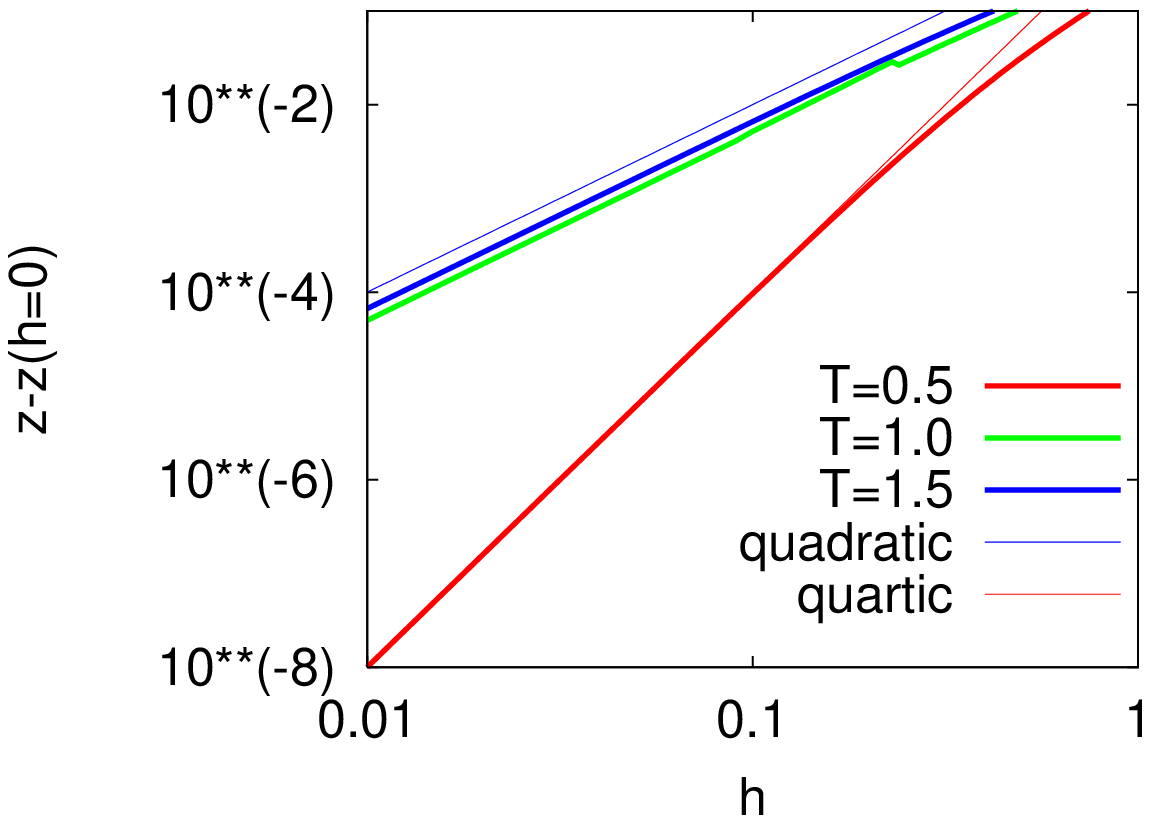}
\end{flushright}
\caption{Left: the Lagrange multiplier $z$ as a function of 
the strength of the applied field $h$. Right: check of the power law
approach to $z(h=0)$ for $h\to 0$ ($h^2$ at high and critical 
temperatures and $h^{2/(1-\alpha)} = h^4$ at low temperatures).}
\label{fig1}
\end{figure}

The magnetisation in the direction of the field,
 equation~(\ref{eq:magn-dir-field}), reads in this case
\begin{displaymath}
m =  \frac{h}{2} \left[ z - \sqrt{z^2-4} \right]
\; .
\end{displaymath}

The non-linear susceptibilities can also be worked out in detail and agree
with the results of the previous section in the case where $c_0 = 1/\pi$.

\subsection{Phenomenology} 

With the aim of testing the relevance of this family of very of simple
models to describe real spin-glass systems, we compare the critical
exponents computed here  with the ones measured experimentally by
L\'evy and Ogielsky~\cite{Leog,Levy}, obtained numerically by a number
of authors~\cite{Kari}, and proposed by Fisher and Huse on the basis
of the droplet model~\cite{Fihu}. In this way we try to find an optimal 
value of $\alpha$ to match the experimental results.

\subsubsection{Comparison with  experimental and numerical results.}

L\'evy and Ogielsky~\cite{Leog,Levy} measured ac non-linear
susceptibilities in a dilute AgMn alloy with the characteristics of a
Heisenberg spin-glass in three dimensions.  In their experimental
studies they identified a finite critical temperature and studied
the critical singularities of the nonlinear susceptibilities 
in the static limit.

Above $T_c$ L\'evy and Ogielsky found
\begin{equation}
\chi_3 \sim \left(\frac{T-T_c}{T_c} \right)^{-\gamma} 
\qquad \mbox{with} \qquad 
\gamma =  2.3 \pm 0.15
\; , 
\label{gamma-Levy}
\end{equation}
and 
\begin{equation}
-\frac{\chi_5}{\chi_3} \sim 
-\frac{\chi_7}{\chi_5} \sim \chi_3^{1+\beta/\gamma}
\qquad \mbox{with} \qquad \beta\sim 0.9\pm 0.2
\; . 
\label{hyperscaling}
\end{equation}
In the spherical disordered models we found
\begin{displaymath}
\chi_3 \sim (T-T_c)^{\frac{-(1-\alpha)}{\alpha}}
\end{displaymath}
so, comparison with  (\ref{gamma-Levy}) implies 
\begin{displaymath}
\alpha \sim 1/3.3 \sim 0.3
\; . 
\end{displaymath}
In addition we found, 
\begin{displaymath}
-\frac{\chi_5}{\chi_3} \sim 
-\frac{\chi_7}{\chi_5} \sim (T-T_c)^{-1/\alpha} \sim 
\chi_3^{1/(1-\alpha)}
\; . 
\end{displaymath}
and thus the  hyperscaling relation (\ref{hyperscaling})
gives
\begin{equation}
\chi_3^{1+\beta/\gamma}
\sim \chi_3^{1/(1-\alpha)}
\qquad \mbox{with} \qquad 
\gamma= \frac{1-\alpha}{\alpha}
\qquad 
\Rightarrow \qquad \beta = 1 
\label{hyperscaling2}
\end{equation}
for all spherical models independently of $\alpha$ for $\alpha \in (0,1)$. 
Note that
$\beta=1$ is consistent with the analytic behaviour of the order
parameter $q_{ea}$ in the low temperature phase.

It is interesting to note that, as summarised in a recent review
article~\cite{Kari}, most isotropic spin-glasses have $\beta \sim
0.9-1.1$ and $\gamma\sim 1.9-2.3$.  The first value is the one we
found for all spherical models, the second one implies $\alpha \sim
0.3$. A scenario including a decoupling of spin and chiral 
order in Heisenberg spin-glasses has been proposed by Kawamura~\cite{Ka}. 

As for Ising spin-glasses, both experiments and simulations point to a
larger value of $\gamma$. Kawashima and Rieger~\cite{Kari} stress that
it is now well established that there is a conventional second order
finite temperature transition with a diverging
$\chi_{SG}$~\cite{Palassini,Ballesteros} and quote, basically, $\beta
\sim 0.5$ and $\gamma\sim 4$ for these `easy-axis' systems.  Daboul,
Chang and Aharony~\cite{Daah} estimated $\gamma$ for the Ising
spin-glass model on a hypercubic lattice in $d\geq 4$ with different
distributions of the coupling strengths using high temperature
expansions. The values they find for higher dimensions also suggest a
rather high $\gamma$ in $d=3$.

We then conclude that, as expected, the spherical model is 
more adequate to describe the high temperature critical 
behaviour of isotropic rather than Ising-like systems.

Below $T_c$ the dynamics are so slow that L\'evy and Ogielsky could
not identify a static limit in zero applied field.  Aging effects come
into play~\cite{Vi} and one has to analyse experimental, as well as
numerical, data very carefully. L\'evy's data in a {\it finite field}
are consistent with $\gamma'=\gamma$, with $\gamma'$ defined from
$\chi_3$ which is finite below $T_c$ in the experimental case. The
value $\alpha\sim 0.3$ that we extracted from the high $T$ analysis
leads, however, to a diverging $\chi_3$ both in the {\it zero applied
field limit} below $T_c$, see equation~(\ref{eq:mh-lowT}) obtained with
(ii) $h\to 0 \;\; N\to \infty$, and in the opposite order of limits, see
equation~(\ref{eq:static-value}), obtained with (i) $N\to\infty \;\; h\to 0$.
Not surprisingly, spherical disordered models cannot capture all
details of real spin-glasses.

\subsubsection{$T<T_c$, comparison with  the droplet model.}

We now compare the critical behaviour of the spherical 
disordered models to that of the droplet theory of spin-glasses~\cite{Fihu}.
We shall distinguish the predictions of the latter 
for Ising and continuous spins.

In the Ising spin-glass phase, $T<T_c$, and in the limit
(ii) $h\to 0$ after the thermodynamic limit $N\to\infty$, 
Fisher and Huse propose
\begin{displaymath}
m(h) \sim h + A h^{d/(d-2\theta)}
\end{displaymath}
while we have 
\begin{displaymath}
m(h) \sim h + A h^{(1+\alpha)/(1-\alpha)}
\; . 
\end{displaymath}
An equivalence between the two implies
\begin{equation}
\alpha = \frac{\theta}{d-\theta}
\; . 
\label{alpha-theta}
\end{equation}
In the reversed order of limits (i) $N\to\infty \;\; h\to 0$, Fisher
and Huse have
\begin{displaymath}
\chi_3 \sim N^{1+\theta(1+\phi)}
\end{displaymath} 
and since we find $\chi_3\sim N$ a comparison leads to 
\begin{equation}
\theta(1+\phi)=0 \qquad \Rightarrow \qquad 
\theta=0 \qquad \mbox{or} \qquad \phi=-1
\; . 
\label{eq:oneoranother}
\end{equation}
Using equation~(\ref{alpha-theta}) the first option, $\theta=0$, yields
$\alpha=0$. Note that $\theta=0$ is usually associated to the replica
symmetry breaking scenario and it has been found numerically 
in $d=2$ for $J_{ij}=\pm 1$~\cite{d2}. 

J\"onsson {\it et al}'s experimental results for the dynamic
relaxation of the AgMn Heisenberg spin-glass compound analysed with
the (slightly modified) droplet scaling imply $\theta\sim
1$~\cite{Jonsson}.  If we still use the relation (\ref{alpha-theta})
and fix $\theta=1$ we then conclude $\alpha=1/2$ (setting
$d=3$). Interestingly enough, $\alpha=1/2$ corresponds to the
spherical SK model but also the spherical ferromagnet in the continuum
limit (the Laplacian in three dimensions leads to a density of states 
approaching the edge with this power).  

\section{Langevin Dynamics}
\label{sec:langevin}

In this section we compute the temporal behaviour of the magnetisation
in the direction of the applied field as a function of time for a system 
quenched from infinite temperature at $t=0$. The dynamics we study is
Langevin dynamics as is the case in most previous studies of the 
dynamics of the spherical SK 
model~\cite{KTJ}-\cite{Becuig}

In the  basis where the matrix $J$ is diagonal the stochastic evolution
equations describing the Langevin dynamics of the system are
\begin{equation}
{\partial s_\mu\over \partial t} = (\mu-z)s_\mu + h \sigma_\mu + \eta_\mu.
\label{lang}
\end{equation}
In the basis of the eigenvalues the  $\sigma_\mu$ are again uncorrelated
and of unit variance. The white noise terms have correlation function
\begin{displaymath}
\langle \, \eta_\mu(t) \eta_{\mu'}(t') \, \rangle 
= 2T \, \delta_{\mu\mu'} \, \delta(t-t').
\end{displaymath}
The term $z$ is a dynamical Lagrange multiplier which enforces the spherical 
constraint and which must be calculated self-consistently. The solution
to equation~(\ref{lang}) is 
\begin{displaymath}
s_\mu(t) = s_\mu(0) \; {\exp(\mu t)\over \Gamma(t)}
+{\exp(\mu t)\over \Gamma(t)} \int_0^t ds \ (h \sigma_\mu + \eta_\mu(s)
) \; {\exp(-\mu s) \, \Gamma(s)}
\end{displaymath}
where $\Gamma(s) \equiv \exp\left(\int_0^t ds\ z(s)\right)$. Now
assuming that the initial conditions are such that they are
uncorrelated with the applied field and also assuming that $\langle
s_\mu(0) s_{\mu'} (0)\rangle = \delta_{\mu\mu'}$ we obtain the
following equation for the magnetisation in the direction of the
applied field:
\begin{displaymath}
m(t) = {1\over N}\sum \sigma_\mu s_\mu(t)  = {h\over\Gamma(t)}
\int_0^t ds\ \langle\langle \, \exp\left[\mu(t-s)\right]\, \rangle\rangle 
\, \Gamma(s)
\end{displaymath}
(the field is applied at the preparation time $t=0$ and 
subsequently kept fixed). 
The self-consistent equation for $\Gamma$ is
\begin{eqnarray}
\Gamma^2(t) &=& \langle\langle \, \exp(2\mu t)\, \rangle\rangle
+ 2T \int_0^t ds\ \langle\langle \, \exp\left[2\mu(t-s)\right] 
\, \rangle\rangle \; \Gamma^2(s)
\nonumber\\
&& 
+ h^2  \int_0^t dsds'\ \langle\langle \, \exp\left[\mu(2t-s-s')\right]
\, \rangle\rangle \; \Gamma(s) \Gamma(s')
\; .
\end{eqnarray}
We restrict out attention to the dynamical
behaviour of just the linear and first non-linear susceptibilities. 
The above equations are thus solved perturbatively to $O(h^3)$ to give
\begin{eqnarray}
m(t) &=& h \int_0^t ds\ \langle \langle \exp\left[\mu(t-s)\right]\,
\rangle\rangle \; {\Gamma_0(s) \over \Gamma_0(t)} 
\nonumber\\
&&
+ h^3\int_0^t ds \
\langle \langle \exp\left[\mu(t-s)\right]\, \rangle\rangle \;
{\Gamma_1(s)\Gamma_0(t) -\Gamma_1(t)\Gamma_0(s) \over \Gamma_0^2(t)}
\; ,
\end{eqnarray}
where $\Gamma_0$ obeys
\begin{equation}
\Gamma^2_0(t) = \langle\langle \, \exp(2\mu t)\, \rangle\rangle
+ 2T \int_0^t ds\ \langle\langle \, \exp\left[2\mu(t-s)\right] 
\, \rangle\rangle \; \Gamma_0^2(s) 
\; , 
\label{eqG0}
\end{equation}
and $\Gamma_1$ is given by
\begin{eqnarray}
2\Gamma_0(t)\Gamma_1(t)
&=& 
  \int_0^t dsds'\ 
\langle\langle \, \exp\left[\mu(2t-s-s')\right]\, \rangle\rangle
\; 
\Gamma_0(s) \Gamma_0(s') 
\nonumber\\
&&
+ 
4T \int_0^t ds\ \langle\langle \, \exp\left[2\mu(t-s)\right]\, \rangle\rangle
\; \Gamma_0(s)\Gamma_1(s)
\; . 
\label{eqG1}
\end{eqnarray}

The dynamical linear  susceptibility is then
\begin{equation}
\chi_1(t) =  
\int_0^t  ds\ \langle \langle \exp\left[\mu(t-s)\right]\, \rangle\rangle 
\;
{\Gamma_0(s) \over \Gamma_0(t)}
\; , 
\label{chi1t}
\end{equation}
and we may write the dynamical nonlinear susceptibility as
\begin{equation}
\chi_3(t) = -6 {\Gamma_1(t)\over \Gamma_0(t)} \; \chi_1(t) +
{6\over \Gamma_0(t)}\int_0^t ds\ 
\langle \langle \exp\left[\mu(t-s)\right]\, \rangle\rangle \; \Gamma_1(s) 
\label{eqchi3a}
\; .
\end{equation}

\subsection{Low temperatures} 

In \cite{cude}  the solution of equation~(\ref{eqG0}) for a general
$\rho(\mu)$ in the ageing regime $T<T_c$ was found, this result can be 
compactly written as
\begin{equation}
\Gamma_0^2(t) = \int d\mu\  q(\mu)\exp(2\mu t) 
\; , \label{eqGrep}
\end{equation}
with $q(\mu)$ given by
\begin{equation}
q(\mu) = {\rho(\mu)\over [1- T \chi(\mu)]^2}\label{eqqm}
\end{equation}
and
\begin{equation}
\chi(\mu) = P\int d\lambda\  {\rho(\lambda)\over \mu -\lambda}
\; ,\label{chimu}
\end{equation}
where $P$ denotes the principal part. For the sake of completeness
 we re-derive the 
result equation~(\ref{eqqm}) in a new more direct way. First if we assume
the representation equation~(\ref{eqGrep}) we find
\begin{eqnarray}
&& 
\int d\mu \ \rho(\mu) \exp(2\mu t)
=
\int d\mu  \ q(\mu) \exp(2\mu t) 
\nonumber\\
&&
\qquad 
+ T \int  d\mu d\mu'\, \rho(\mu') \, q(\mu)
\left[ \exp(2\mu' t) - \exp(2\mu t)\over (\mu'-\mu)\right]
\; .
\end{eqnarray}
Notice that the apparent singularity in the second integral on the 
right hand side at $\mu=\mu'$ is not really present and we can replace the
integral by its principal part. Equating the coefficients of $\exp(2\mu t)$ 
in the above equation now yields 
\begin{equation}
q(\mu) = \rho(\mu) + 
T \, \chi(\mu) \, q(\mu) + T \, \rho(\mu)P \int d\mu' \; {q(\mu')
\over \mu-\mu'}
\; .
\label{eqqi}
\end{equation}
The Laplace transform of equation~(\ref{eqG0}) reads
\begin{displaymath}
\widetilde{\Gamma_0^2}(p) = {\int d\mu \  {\rho(\mu)\over p -2\mu}\over
1- 2T \int d\mu \ {\rho(\mu)\over p -2\mu}},
\end{displaymath}
where
\begin{eqnarray}
\widetilde{\Gamma_0^2}(p) &\equiv& \int_0^\infty dt \ \exp(-pt) 
\Gamma_0^2(t)
=\int d\mu \; {q(\mu) \over p -2\mu}
\; .
\end{eqnarray} 
The  above now implies that
\begin{eqnarray}
P\int d\mu' \; {q(\mu') \over \mu -\mu'} 
&=& 
{P\int d\mu'\ {\rho(\mu')\over \mu -\mu'}\over
1- T P\int d\mu'\ {\rho(\mu')\over \mu -\mu'}}
= {\chi(\mu)\over 1-T \chi(\mu)}
\; .
\end{eqnarray}
Using this result for the last term in equation~(\ref{eqqi}) we obtain 
equation~(\ref{eqqm}).

\subsection{The Gamma function}

Let us analyse the asymptotic behaviour of $\Gamma$ in the two limits.
At late times the dominant contribution
to $\Gamma_0^2(t)$ comes from around $\mu = \mu_0$. Expanding about this point
we find 
\begin{displaymath}
 \Gamma_0^2(t)\approx\int_0^\infty d\epsilon\
{c_0\epsilon^{\alpha}\over( 1- T\chi(\mu_0))^2} \exp[2(\mu_0-\epsilon)
t],
\end{displaymath}
where we have used equation~(\ref{eqexrho}) for the density of states
$\rho$ at the edge of the spectrum. 

In what follows without loss of generality we restrict ourselves to
the case where $\mu_0 =0$ which can be achieved by a constant shift in
the energy by using the interaction matrix $J'= J-\mu_0 I$.
From the definition of $\chi(\mu)$
in equation~(\ref{chimu}) and equation~(\ref{eqbc}) we find 
\begin{equation}
\Gamma_0^2(t)\approx {c_0 \Gamma(1+\alpha)\over (2t)^{1+\alpha}( 1-
{T\over T_c})^2 },\label{g0as}
\end{equation}
where $\Gamma$ in the above is the standard gamma function defined as
\begin{displaymath}
\Gamma(x) =\int_0^\infty dt \ t^{x-1} \exp(-t).
\end{displaymath}

We now compute the large time behaviour of $\Gamma $ in the 
region $T<T_c$. Notice that the Laplace transform of $\Gamma_0$ at small
$p$ is dominated by the large $t$ behaviour of $\Gamma_0(t)$, 
for $p$ small and $\alpha <1$  we have that
\begin{equation}
\widetilde{\Gamma_0}(p)\approx A\int_0^\infty dt \exp(-pt)
t^{-\frac{1+\alpha}{2}}
\approx 
A \; \Gamma\left({1-\alpha\over 2}\right)p^{\alpha -1
\over 2},\label{ltg0}
\end{equation} 
where
$
A = \sqrt{c_0 \Gamma(1+\alpha)}
2^{-(1+\alpha)/2}
(1- T/T_c)
$.

In the case in which we keep the $N$ dependence, useful to study
case (i), we find
\begin{equation}
\widetilde{\Gamma_0}(p) \sim \frac1{\sqrt{N (1-T/T_c)}} \; 
\frac1{p-\frac{T}{N(1-T/T_c)}}
\; . 
\label{eq:Gamma0p}
\end{equation}

\subsection{The linear susceptibility}

Equation~(\ref{chi1t}) can now be written as
\begin{displaymath}
\chi_1(t) = {K(t)\over \Gamma_0(t)}
\; ,
\end{displaymath}
with
\begin{displaymath}
K(t) = \int_0^t ds \; 
\langle\langle \, \exp\left[\mu(t-s)\right] \rangle\rangle
\; 
\Gamma_0(s) \; .
\end{displaymath}
The Laplace transform of $K$ is given by
\begin{displaymath}
{\tilde K}(p) = \langle \langle {1\over p -\mu}\rangle \rangle 
\; \widetilde{\Gamma_0}(p)
\; .
\end{displaymath}
We may thus write
\begin{eqnarray}
{\tilde K}(p) 
&=& 
\left[\langle \langle -{1\over\mu}\rangle \rangle
+ p \; 
\langle \langle {1\over\mu(p-\mu)}\rangle \rangle\right]\widetilde{\Gamma_0}(p)
\nonumber \\
&=& \left[{1\over T_c}
+ p \; 
\langle \langle {1\over\mu(p-\mu)}\rangle \rangle\right]
\widetilde{\Gamma_0}(p) 
\; .
\label{eq:Kp}
\end{eqnarray}

\vspace{0.25cm}
\noindent
{(ii) $h\to 0 \;\; N\to\infty$}
\vspace{0.25cm}

The term $\langle \langle 1/(\mu(p-\mu))\rangle \rangle$ 
diverges as $p\to 0$ for $\alpha <1$; the small $p$ behaviour is thus dominated
by the region around $\mu =0$. This gives for small $p$
\begin{eqnarray}
\langle \langle {1\over\mu(p-\mu)}\rangle \rangle &\approx&
-c_0\int_0^\infty d\epsilon {\epsilon^\alpha\over (p+\epsilon)\epsilon}
=-c_0 p^{\alpha-1} I(\alpha)
\; ,
\label{intas}
\end{eqnarray}
where $I(\alpha)$ is as defined by equation~(\ref{ialpha}).
From the above and equation~(\ref{ltg0}) we thus find that for small $p$
\begin{displaymath}
{\tilde K}(p) \approx {\widetilde{\Gamma_0}(p)\over T_c}-c_0 I(\alpha)
A \Gamma({1-\alpha\over 2}) p^{3\alpha -1\over 2} 
\; .
\end{displaymath}
Asymptotically inverting the Laplace transform we obtain
\begin{displaymath}
K(t) \approx {\Gamma_0(t)\over T_c} - {c_0 I(\alpha) A
\Gamma({1-\alpha\over 2}) \over \Gamma({1-3\alpha \over 2}) t^{3\alpha
+1\over 2}}
\end{displaymath}
which finally yields for large $t$
\begin{equation}
\chi_1(t) \approx {1\over T_c} - {c_0 I(\alpha)
\Gamma({1-\alpha\over 2}) \over \Gamma({1-3\alpha \over 2})}
\; t^{-\alpha}
\; . \label{chi1f}
\end{equation}
We thus see that $\chi_1(t)$ decays to its  low temperature
equilibrium value with a power law $t^{-\alpha}$. Interestingly the 
coefficient of this decaying term is negative for $\alpha <1/3$, 
meaning that $\chi_1(t)$  achieves its equilibrium value from below, whereas 
for $\alpha >1/3$ the coefficient is positive and thus $\chi_1(t)$  
achieves its equilibrium value from above.

\vspace{0.25cm}
\noindent
{(i) $N\to\infty \;\; h\to 0$}
\vspace{0.25cm}

In this limit one can analyse $\widetilde{K}(p)$ in (\ref{eq:Kp}) 
keeping the $1/N$ contributions. One finds
\begin{displaymath}
\chi_1(t) = \frac1{T} - \left( \frac1{T} - \frac1{T_c} \right) \;
e^{-\frac{t}{N (\beta-\beta_c)}}
\; . 
\end{displaymath}
Note that if we set $t/N \ll 1$ we recover 
$\chi_1\sim \beta_c$. 

\subsection{The non-linear susceptibility}

We now turn to the results for the nonlinear susceptibility $\chi_3$.
If we define $Q(t) = \Gamma_0(t)\Gamma_1(t)$, from equation~(\ref{eqG1}) 
we find that the Laplace transform of $Q$ obeys
\begin{displaymath}
\widetilde{Q}(p) = 
{\widetilde{f}(p)\over 1-2T \langle\langle \, {1\over p-2\mu}\, \rangle\rangle}
\; ,
\end{displaymath}
where
\begin{displaymath}
f(t) = {1\over 2}\int_0^t dsds'\ \langle \langle 
\exp\left[2\mu t -\mu s-\mu s'\right]\, \rangle\rangle
\; 
\Gamma_0(s)\Gamma_0(s')
\; . 
\end{displaymath}

\vspace{0.25cm}
\noindent
{(ii) $h\to 0 \;\; N\to\infty$}
\vspace{0.25cm}

Let us now focus on this order of limits. 
Making the substitution $w=-\mu t$ and $s=vt$ we find
\begin{displaymath}
f(t) = {1\over 2}\int_0^{-\mu^* t}\rho\left({-w\over t} \right) {dw\over t}
\left[ \int_0^1 t dv \ \exp\left[-w(1-v)\right] \; \Gamma_0(vt)\right]^2.
\end{displaymath}  
We now use the asymptotic form of $\Gamma_0$  
in equation~(\ref{g0as}) to find, for large $t$, 
\begin{displaymath}
f(t)  \approx {A^2 c_0\over 2 t^{2\alpha}}\int dw \; w^{\alpha}\left[
\int_0^1 dv \; {\exp\left[-w(1-v)\right]\over v^{1+\alpha\over 2}}\right]^2
\; . 
\end{displaymath}
Thus $f(t) = B'/t^{2\alpha}$ for large $t$ and one may verify that the constant
$B'$ is finite for $\alpha <1$. Consequently, we obtain 
\begin{displaymath}
\Gamma_1(t) = B \;  t^{\frac12 (1-3\alpha)}
\; .
\end{displaymath} 
The small $p$ behaviour of the Laplace transform of $\Gamma_1$ is thus given by
\begin{displaymath}
\widetilde{\Gamma_1}(p) \approx B \; \Gamma\left({3\over 2}(1-\alpha)\right)
p^{-{3\over 2}(1-\alpha)}
\; .
\end{displaymath}
We now rearrange the result  in equation~(\ref{eqchi3a}) as
\begin{displaymath}
\chi_3(t) = -6 \; {\Gamma_1(t)\over \Gamma_0(t)}\left[\chi_1(t)-{1\over
T_c}\right] - 6 \; {L(t)\over \Gamma_0(t)}
\end{displaymath}
where 
\begin{displaymath}
L(t) = {\Gamma_1(t) \over T_c} -\int_0^t ds\ \langle\langle \, 
\exp\left[\mu(t-s)\right]\, \rangle\rangle \; \Gamma_1(s)
\; . 
\end{displaymath}
The Laplace transform of $L$ is given by
\begin{eqnarray}
{\tilde L}(p) = \left( {1\over T_c} -\langle
\langle {1\over p-\mu}\, \rangle\rangle \right) 
\;  
\widetilde{\Gamma_1}(p)
= - p \; \widetilde{\Gamma_1}(p)\langle \langle {1\over \mu (p-\mu)}\, 
\rangle\rangle
\; . 
\end{eqnarray} 
For small $p$ using the asymptotic result for $\Gamma_1$ and equation~(\ref{intas})
we find
\begin{displaymath}
{\tilde L}(p) \approx c_0 \, B \, I(\alpha) \; 
\Gamma\left({3\over 2}(1-\alpha)\right) p^{{5\over 2}\alpha -{3\over 2}}
\; ,
\end{displaymath}
which implies that the  late time behaviour of $L$ is 
\begin{displaymath}
L(t) \approx  {c_0 \, B \, I(\alpha) \; \Gamma\left({3\over 2}(1-\alpha)\right)\over \Gamma({3\over 2}-{5\over 2}\alpha) \; 
t^{{5\over 2}\alpha -{1\over 2}}}
\; .
\end{displaymath}
We can compute the other contribution to $\chi_3$ using the asymptotic
result for $\chi_1$ equation~(\ref{chi1f}) to obtain the result
\begin{displaymath}
\chi_3(t) \approx -{6 \, c_0 \, B \, I(\alpha) \over A}\left[{\Gamma\left({3\over 2}
(1-\alpha)\right)\over \Gamma({3\over 2}-{5\over 2}\alpha)}-
{\Gamma({1-\alpha\over 2}) \over \Gamma({1-3\alpha \over 2})}\right]
t^{1-2\alpha}.
\end{displaymath}
We therefore see that when $\alpha < 1/2$ the nonlinear susceptibility
diverges as $\chi_3(t) \approx C t^{1-2\alpha}$. When $\alpha >1/2$ 
$\chi_3(t)$ decays to zero as $1/t^{2\alpha-1}$. These dynamical results
are of course in agreement with the static calculations carried out earlier.
The coefficient of this term is determined by the sign of that in the 
square brackets, the first prefactor being negative. Numerical evaluation
of the factor in square brackets confirms that it is positive for $\alpha
\in (0,1)$ and thus the dynamical nonlinear susceptibility is negative. 
Indeed for $T>T_c$ the static value is negative and the dynamical calculation
confirms the divergence to an infinite negative value for $\alpha < 1/2$.
  
\vspace{0.25cm}
\noindent
{(i) $N\to \infty \;\; h \to 0$}
\vspace{0.25cm}

 In order to 
compute $\chi_3(t)$ we need to compute $\Gamma_0(t)$ and $\Gamma_1(t)$  
keeping the correction to the leading exponential in time terms. 
From equations~ (\ref{eqG1}) and (\ref{eq:Gamma0p}) we find 
\begin{eqnarray}
\Gamma_0(t) &\sim& \frac1{\sqrt{N \left(1-T/T_c \right)}}
\; 
e^{\frac{t}{N (\beta-\beta_c)}}
\; , 
\\
\Gamma_1(t) &\sim& \frac1{2T^2} \sqrt{N \left(1-T/T_c\right)}
\; \left\{1+2t/[N(\beta-\beta_c)]\right\}
\; e^{\frac{t}{N (\beta-\beta_c)}}
\; . 
\end{eqnarray} 
Using equation~(\ref{eqchi3a}) it can be verified that the asymptotic limit
of $\chi_3(t)$ is the static value (\ref{eq:static-value}) up to a
correction term that decays exponentially as $e^{-\frac{t}{N
(\beta-\beta_c)}}$.

\section{Discussion and Conclusions}
\label{sec:discussion}
We have studied in detail the linear and first non-linear
susceptibilities of generalised random orthogonal model spherical spin
glasses. Their physics is completely determined by the density of
states of the two body interaction matrix. In particular the
exponent, $\alpha$ describing how the density of states vanishes at
the upper edge, determines completely the critical behaviour at the
phase transition and the dynamical evolution of $\chi_1(t)$ and
$\chi_3(t)$ in the limit (ii) considered here where the limit $h\to 0$
is taken after the limit $N\to \infty$ in the computations.  With
respect to the Gaussian $p=2$ spin glass model we see that the
existence of the parameter $\alpha$ gives us the possibility to carry
out a more meaningful comparison between the model and experimental
and droplet scaling theories. An interesting aspect of this work is
that it clearly demonstrates the possibility that $\chi_3$ may appear
to be finite for certain classes of model (with $\alpha \geq 1/2$) if
the applied fields used to carry out the measurements place us in the
regime of limits (ii). However it is in this same region where
the cusp in the linear susceptibility exists. These results, though
for a somewhat idealised mean field model, could well have some
bearing on the interpretation of susceptibility and magnetisation
measurements in disordered and frustrated spin systems~\cite{frustrated}. 
Finally we
have mentioned that the models considered here will still exhibit a
finite temperature transition in the presence of an external magnetic
field if $\alpha >1$. In this case it is the higher oder ($n>3$)
susceptibilities which will diverge and it will be interesting to
study, in particular, the dynamics of these models.


\ack 

LFC and DSD acknowledge hospitality and
financial support from the Isaac Newton Institute, at the University
of Cambridge, UK and the KITP, University of California at Santa
Barbara, USA where part of this work was carried out.  This research was
supported in part by NSF grant No PHY99-07949 (LFC and DSD). HY
acknowledges support from the Japanese Society of Promotion of Science
and CNRS and hospitality from the Laboratoire de Physique Th\'eorique
et Hautes Energies at Jussieu, Paris, France.  LFC and DSD are  members of the
Institut Universitaire de France.

\end{document}